# Giant anisotropy of the magnetocaloric effect in the orthovanadate TbVO$_4$ single crystals

M. Balli [1, *], S. Mansouri [2], D. Z. Dimitrov [3, 4, 5], P. Fournier [2, 6], S. Jandl [2], Jenh-Yih Juang [5]

[1] AMEEC Team, LERMA, College of Engineering &Architecture, International University of Rabat, Parc Technopolis, Rocade de Rabat-Salé, 11100, Morocco.

[2] Institut Quantique, Université de Sherbrooke, J1K 2R1, QC, Canada.

[3] Institute of Solid State Physics, Bulgarian Academy of Science, Sofia 1184, Bulgaria.

[4] Institute of Optical Materials and Technologies, Bulgarian Academy of Sciences, Sofia 1113, Bulgaria

[5] Department of Electrophysics, National Chiao Tung University, 1001 Ta Hsueh Road, Hsinchu, Taiwan.

[6] Canadian Institute for Advanced Research, Toronto, Ontario M5G 1Z8, Canada.

**Abstract**

It is known that the Zircon-type orthovanadates RVO$_4$ show promise in many different applications as catalysts and optical materials. In this work, we demonstrate that the TbVO$_4$ compound can be also used as magnetic refrigerant in efficient and ecofriendly cryocoolers due to its strong magnetocaloric effect at low temperature regime. The application of a relatively low magnetic field of 2 T along the easy magnetization axis (a) gives rise to a maximum entropy change of about 20 J/kg K at 4 K. More interestingly, under sufficiently high magnetic fields, the isothermal entropy change -$\Delta S_T$ remains approximately constant over a wide temperature range which is highly appreciated from a practical point of view. In the magnetic field change of 7 T, -$\Delta S_T$ that reaches roughly 22 J/kg K remains practically unchanged between 0 and 34 K leading to an outstanding refrigerant capacity of about 823 J/kg. On the other hand, the lowering of crystallographic symmetry from the tetragonal to the orthorhombic structure occurring close to 33 K as confirmed by Raman scattering data results in a strong magnetic anisotropy. Accordingly, strong thermal effects can be also obtained simply by spinning the TbVO$_4$ single crystals between their hard and easy orientations in constant magnetic fields instead the standard magnetization-demagnetization process. Such rotating magnetocaloric effects would open the way for the implementation of TbVO$_4$ in a new generation of compact and simplified magnetic refrigerators that can be dedicated to the liquefaction of hydrogen and helium.

*Mohamed.balli@uir.ac.ma

# I. Introduction

In our modern society, the search for new advanced materials that exhibit outstanding functional properties becomes primordial to deal with numerous environmental and economic issues such as the global warming and the resources energy scarcity. In this context, the implementation of magnetocaloric materials that can be used as active refrigerants in cooling (heating) systems would greatly contribute to reduce greenhouse gas (GHG) emissions while enabling efficient refrigerators [1-6]. Based on the magnetocaloric effect (MCE) that can be defined as thermal changes experienced by some magnetic materials under the variation of applied external magnetic fields, magnetic cooling technique would allow us greater energy savings due to its potential higher thermodynamic efficiency [1-6]. However, the development and the discovery of new performant magnetocaloric compounds is a key parameter for the commercialization of this promising technology. Aiming to replace rare earth elements-based materials which are widely used in room-temperature magnetic refrigeration demonstrators, worldwide research activities on magnetocalorics were conducted during the last two decades leading to a wider variety of interesting materials from both economical and performance points of view. This includes for example free (reduced) rare-earth compounds such as MnFe(P, Si, Ge, As) and $LaFe_{13-x}Si_x$ intermetallics [1, 3, 4, 6]. However, despite of their large entropy changes under magnetic fields reachable by permanent magnets, the adiabatic temperature change ($\Delta T_{ad}$) of these reported compounds fails to exceed its equivalent exhibited by the benchmark gadolinium metal [1]. In addition, the moderate chemical stability and the mechanical brittleness would restrict their utilization as refrigerants in commercial devices, opening the avenue for further investigations in the field [1]. In this way, the $(La, Sr, Pr)_{2/3}Ca_{1/3}MnO_3$ -based oxides could be of interest in magnetic cooling on account particularly to their strong chemical stability [1, 5]. Besides, the magnetocaloric properties of such materials must be markedly improved before their implementation in functional devices. For this purpose, a high fundamental understanding of the coupling between their structural, magnetic and electrical degrees of freedom with the aim to approach the MCE theoretical limit may open the door to their practical application.

Magnetocaloric materials at low temperature regime are also of great importance in some critical applications such as gas liquefaction, space industry and scientific instruments. Particularly, the liquefaction as well as the safe storage of hydrogen that is considered as one of the most promising energetic alternatives is a great issue that merits intensive investigations [7]. In this context, an innovative hydrogen liquefier based on the active magnetic refrigeration (AMR) approach [2, 3] was reported in Matsumoto et al [8] and Numazawa et al [9]. The latter have used the magnetocaloric garnet $Dy_{2.4}Gd_{0.6}Al_5O_{12}$ as refrigerant, resulting in the decrease of the hydrogen temperature at the cold side down to 20 K. However, the large specific heat usually shown by the garnet compounds lower drastically their MCE in term of the adiabatic temperature change that reaches only 1 to 2 K in the magnetic field change of 1 T [8, 9]. In order to deal with these drawbacks, interesting materials with better magnetocaloric performance in the temperature range below 30 K were recently reported including both intermetallics [10-13] and oxides such as $RMO_3$ (M = metal) [14-21]. On the other hand, the $RMn_2O_5$ multiferroics are also of great importance in the cryogenic temperature range not only because of their strong MCEs that are usually obtained by varying magnetic fields along their easy-axes, but also due to their strong magnetocrystalline anisotropy. The latter enables additional thermal effects that can be obtained by simply rotating their crystals in constant magnetic fields, opening the way for more compact and simplified efficient cryocoolers [22, 23]. However, in addition to long term mechanical and chemical stabilities, the potential magnetocaloric candidates must exhibit sufficiently high MCEs over a wide temperature range. Such a behavior would render the refrigeration process more efficient under the AMR and Ericsson cycles [1-3, 10]. Keeping in mind these requirements, the magnetocaloric features of $TbVO_4$ are investigated aiming to gain insight on its applicability as an efficient refrigerant at the cryogenic temperature range.

The $TbVO_4$ compound belongs to the $RVO_4$ (R = rare earth) family of orthovanadates which is still attracting worldwide attention due particularly to its fascinating structural and electronic properties under intense magnetic fields [24-29] as well as their potential integration in optical devices as polarizers and fluorescent/laser materials [30]. They usually undergo a tetragonal zircon type symmetry with a space group $I4_1/amd$ at room temperature in which the vanadium and rare earth metals are surrounded by oxygen tetrahedrons ($VO_4$) and bisdisphenoids ($RO_8$) (see Fig. 1), respectively [24-31]. Some of them unveils phase transitions below 40 K lowering the crystallographic symmetry to the orthorhombic structure. Such as a transformation is attributed to a cooperative Jahn-Teller (JT) effect arising from the strong coupling between the crystalline lattice and the electronic order of $R^{3+}$ ions [24-29]. On the other hand, the $R^{3+}$ ions which are located at magnetically equivalent sites unveil a weak magnetic order even at very low temperatures. In fact, the vast majority of $RVO_4$ usually experiences a magnetic order below 4 K. For the $TbVO_4$, an antiferromagnetic ordering takes place close to 0.6 K [26]. It is worthy to underline that similarly to their optical properties, a great interest was also given to the magnetic behavior of $RVO_4$ orthovanadates including the $TbVO_4$ compound [24-38]. In this paper, we particularly investigate the magnetic and magnetocaloric properties of $TbVO_4$ crystals with the aim to enlarge our understanding of the fundamental mechanisms behind some of the $RVO_4$ functional properties and to figure out whether $TbVO_4$ is suitable for application in cryogenic magnetic cooling. With regards to the latter point, our findings demonstrate that the here studied compound presents both outstanding conventional and rotating magnetocaloric features, opening the door for its implementation in a new generation of magnetocaloric devices. Also, the impact of crystal electrical field on the magnetism and resulting magnetocaloric features in TbVO$_4$ is deeply analyzed.

## II. Experimental

TbVO$_4$ single crystals have been grown by high temperature solution technique as well described in Ref. 39. As a first step, polycrystalline TbVO$_4$ was synthesized by the solid-state reaction method using stoichiometric ratios of Tb$_2$O$_3$ and V$_2$O$_5$ with minimum purity of 99.99%. The starting oxides were ground, mixed, compacted, and then calcinated in oxygen atmosphere at 650 °C for 48 h. The reacted product was ground and mixed with V$_2$O$_5$ flux in V$_2$O$_5$: TbVO$_4$ = 7:1 ratio. The mixture was melted and heated to 1050 °C for 48 h in a platinum crucible of 50 mm diameter and 60 mm depth, covered with a platinum lid. Single crystals were obtained by cooling the solution from 1050 to 700 °C at a cooling rate of 1 °C/h. The residual flux was separated from the as-grown crystals by decanting. The obtained crystals remained on the bottom and the walls of the crucible. They present a high optical quality with tetragonal rectangular shape and typical size of 4 x 3 x 0.5 mm. The quality of crystalline structure as well as the symmetry of the crystals investigated here are checked by using Raman scattering investigations. Micro-Raman spectra were collected in the temperature range going from 5 to 300 K with the help of a Labram-800 spectrometer equipped with a microscope, a He-Ne laser and a nitrogen-cooled charge coupled device (CCD) detector. The power of the laser beam was maintained under 0.3 mW which enables us to avoid local heating of the sample. A careful analysis of Raman spectra reported in Figure 2 unveils an outstanding crystalline quality while the observed signatures at room-temperature confirm a tetragonal symmetry with space group I4$_1$/amd. The latter is lowered as expected to an orthorhombic structure with Fddd space group at temperatures below 30 K as clearly shown in Figure 2. Infrared transmission spectra of TbVO$_4$ as a function of temperature with a resolution of 1 cm$^{-1}$ were collected by using a Fourier transform interferometer BOMEM DA3.002 equipped with a quartz-halogen source, a CaF2 beamsplitter and an InSb detector. Magnetization measurements as a function temperature and magnetic field were carried out by using a superconducting quantum interference device (SQUID) magnetometer from Quantum Design (MPMS XL).

## III. Results and discussion

The temperature dependence of zero-field cooled (ZFC) and field-cooled (FC) magnetization is reported in Figure 3 along the a and c axes for the TbVO$_4$ single crystal. As shown, the nearly reversible behavior of magnetization as a function of temperature reveals the absence of thermal hysteretic effects in this orthovanadate oxide. Additionally, as the temperature decreases, a rapid magnetization change can be clearly seen around 33 K (inset of Figure 3). The observed feature can be ascribed to the occurrence of a crystallographic transformation from the high-temperature tetragonal structure to the low-temperature orthorhombic symmetry arising from a B$_{2g}$ strain [40] as confirmed by the reported Raman scattering data in Figure 2. In fact, the TbVO$_4$ compound experiences a cooperative Jahn-Teller distortion at T$_Q$ = 33.1 K that leads to a strong change in the electronic energy levels as reported in Ref 26. The reducing of the symmetry from tetragonal to orthorhombic structure is evidently observed through the splitting of the E$_g$ mode at 826 cm$^{-1}$ as shown in Figure 2(b). Upon further lowering the temperature down to 2 K, the magnetization following the main crystallographic orientations increases rapidly without exhibiting any significant anomaly. This is mainly attributed to the thermal population of the crystal field (CF) levels [32]. In fact, the magnetic behaviour of the pseudo-Ising TbVO$_4$ system is related to the strong impact of crystal-field effects on the Tb$^{3+}$ ions electronic energy levels. At the temperature above that of Jahn-Teller distortion, the crystal field splits the lowest ground state into two singlets and a non-Kramers doublet [26, 32-38]. In the temperature range below T$_Q$, a marked change in the energy levels caused by JT interactions have been observed [26, 32-38]. Consequently, the doublets are split into 2 singlets leading to four energy levels where the two lowest singlets are treated as a non-Kramers doublet with S = 1/2 [26, 32-38].

In Figure 4-a we report the inverse magnetic susceptibility (1/χ) as a function of temperature which follows the Curie-Weiss law 1/χ = (T-T$_\Theta$)/C (C is the Curie constant) at high temperatures. Its linear fit leads to paramagnetic curie temperatures (T$_\Theta$) of 14.7 and -70 K along the a and c axes, respectively. This indicates on weak ferromagnetic interactions along the a axis and strong antiferromagnetic couplings along the c direction. The deduced experimental effective magnetic moments from the reverse susceptibility at high temperatures are found to be 9.98 μ$_B$ for the c axis and 9.55 μ$_B$ for the a axis, being closer to the effective magnetic moment of Tb$^{3+}$ given by μ$_{eff}$(Tb$^{3+}$) = 9.72 μ$_B$ [41]. Since the valence shell of V$^{5+}$ ions is empty, the resulting effective magnetic moment is assumed to be zero. On the other hand, as shown in Figure 3, a large deviation between the thermomagnetic curves following the c and a orientations can be particularly seen at low temperatures unveiling the presence of a strong magnetocrystalline anisotropy in TbVO$_4$ crystals as a consequence of the strong spin-orbit coupling of Tb$^{3+}$ magnetic moments. This is clearly supported by the reported magnetic isotherms in Figure 4-b at 2 K for magnetic fields parallel to the a and c orientations. When a 5 T-magnetic field direction is changed from the a axis to the c axis, the magnetization is roughly reduced by 80 % revealing the hard and easy magnetization directions along the c axis and the a axis, respectively. However, it is worthy to mention that magnetization measurements were also performed within the ab-plane following several directions as shown in the inset of Figure 4-b unveiling a quasi-isotropic behaviour which contrast to Ergun et al findings [35]. This point will be deeply investigated in the future.

According to Figure 4-b, the magnetization along the c axis varies slightly when increasing magnetic field following an almost linear trend to reach only 2 $\mu_B$ under a high magnetic field of 5 T. In contrast, along the a axis, the magnetization increases rapidly with magnetic field to saturate after overpassing a relatively low field of about 1.3 T. At the saturation state, the obtained magnetization is 9.07 $\mu_B$ being practically equal to the $Tb^{3+}$ free magnetic moment (9 $\mu_B$) [41]. Since the $V^{3+}$ sublattice does not contributes to the whole magnetization, this result indicates that the magnetic moments of $Tb^{3+}$ sublattice can be completely aligned when applying moderate magnetic fields along the a axis. Accordingly, large magnetocaloric effects are expected in the $TbVO_4$ orthovanadate under magnetic fields that can be reached via permanent magnets which is of great importance from a practical point of view.

One of this paper main scopes concerns the investigation of the magnetocaloric properties in $TbVO_4$ crystals. The MCE is mainly characterized in terms of two principal thermodynamic properties which are the isothermal entropy change ($\Delta S_T$) and the adiabatic temperature change ($\Delta T_{ad}$). Both quantities can be deduced simultaneously from specific heat measurements. Otherwise, $\Delta T_{ad}$ can be directly measured by a system of thermocouples while $\Delta S_T$ can be indirectly deduced from magnetization data using the well-known Maxwell relation given by:

$$\Delta S_T(T, 0-H) = \mu_0 \int_0^H \left(\frac{\partial M}{\partial T}\right)_{H'} dH' \qquad (1)$$

In order to avoid the experimental difficulties usually associated with the calorimetric techniques, the magnetization method which is a simpler technique is widely used to evaluate the MCE in term of $\Delta S_T$. In the absence of hysteresis effects [42] which is the case of the crystals explored here, the Maxwell equation enables us a first and rapid assessment of magnetocaloric materials regarding their applicability in magnetic refrigeration. For this purpose, magnetic isotherms must be carried out in the desired temperature range as reported in Figures 5-a, b for the $TbVO_4$ single crystal. These measured curves are integrated using the numerical form of equation 1 to finally obtain $\Delta S_T$ data. The temperature dependence of $\Delta S_T$ is shown in Figures 6-a, b for $TbVO_4$ under some selected magnetic field variations applied along the c and a axes. As can be seen, the $TbVO_4$ reveals large isothermal entropy changes for magnetic fields higher than 2 T applied along the a axis particularly at low temperatures. In the magnetic field change of 2 T, $-\Delta S_T$ reaches a maximum value of about 22 J/kg K which outperforms that found in some reference cryogenic magnetocaloric oxides containing Dy and Ho elements. For example, under a magnetic field changing from 0 to 2 T along their easy directions in a similar temperature range, the maximum $-\Delta S_T$ values are reported to be only about 3.86 J/kg K for h-HoMnO$_3$ (h = hexagonal) [43], 8 J/kg K for h-DyMnO$_3$ [44], 3 J/kg K for HoMn$_2$O$_5$ [22] and 4 J/kg K for o-DyMnO$_3$ (o = orthorhombic) [45]. The enhancement of the isothermal entropy change in the TbVO$_4$ single crystal following the a axis can be partly attributed to the possibility to fully order the Tb$^{3+}$ magnetic moments by applying low magnetic fields along this direction as shown in figures 4 and 5. Unlike the manganite compounds such as RMnO$_3$ and RMn$_2$O$_5$ [14-23], the absence of highly frustrated magnetism would contribute to this enhancement. On the other hand, the TbVO$_4$ crystals unveil isothermal entropy changes (under 2 T) larger than those reported in a similar family of orthovanadates such as GdVO$_4$ (about 10 J/kg K for 2 T) [31, 33]. More interestingly, the here investigated compound reveals a distinguished feature of $-\Delta S_T$ (T) profile, particularly under sufficiently high magnetic fields. As shown in figure 6-a, the operating magnetocaloric temperature range markedly increases when increasing the magnetic field. For example, in the magnetic field change of 7 T, $-\Delta S_T$ that reaches about 22 J/kg K remains approximately unchanged over the temperature range going from 34 K down to "0 K". Such a behavior is highly appreciated from a practical point view since it opens the way for the implementation of TbVO$_4$ in more efficient Ericsson and AMR thermodynamic cycles [1-3, 10]. Usually, a constant isothermal entropy change can be obtained by combining several materials presenting transition temperatures distributed over the desired temperature range. However, this would generate supplemental costs and engineering challenges while similar performance can be obtained by simply using a single TbVO$_4$ material as shown in Figure 6.

The nearly constant isothermal entropy change over a wide temperature range shown by TbVO$_4$ crystals means that an outstanding refrigerant capacity (RC) is expected in this compound. If the $\Delta S_T$ measures the cooling energy that can be provided by the magnetocaloric refrigerant in an isothermal process, the RC is an important practical parameter that enables us to figure out whether the considered candidates are appropriate for working as refrigerants in functional devices. Indeed, RC evaluates the amount of energy that can be transferred between the hot and cold ends in an AMR cycle. In addition to the magnitude of $\Delta S_T$, RC takes also into account the working temperature range. It is given by

$$RC = \int_{T_C}^{T_H} \Delta S(T) dT \qquad (2)$$

where $T_C$ and $T_H$, are the temperatures at half maximum of $-\Delta S_T$ (T) profile. In the magnetic field variation of 7 T applied along the a axis, the TbVO$_4$ provides a record refrigerant capacity of about 823 J/kg when compared to the most promising magnetocaloric oxides operating in a similar temperature range such as RMO$_3$ and RMn$_2$O$_5$ multiferroics [14]. Considering the latter, the largest RC under 7 T is presented by the orthorhombic DyMnO$_3$ (452 J/kg) [45] and TbMn$_2$O$_5$ (480 J/kg) [23] compounds along their easy magnetization axes. Also, the presented RC by TbVO$_4$ largely overpasses that shown by the best orthovanadates such as HoVO$_3$ (620 J/kg under 7 T) [46] and GdVO$_4$ (360 J/kg under 7 T) [33], phosphates such as GdPO$_4$ (465 J/kg under 7 T) [47], metal-organic Gd(HCOO)$_3$ (390 J/kg under 7 T) [48] and titanates such as EuTiO$_3$ (500 J/kg under 7 T) [20]. For comparison, the RC of some representative cryo-magnetocaloric oxides is reported in Figure 7. As shown, the TbVO$_4$ refrigerant capacity is about twice larger when compared to GdVO$_4$, GdPO$_4$ and Gd(HCOO)$_3$ compounds despite of their "gigantic" maximum $-\Delta S_T$ that reaches 48, 63 and 55 J/kg K under 7 T, respectively. Once again, the giant refrigerant capacity shown by TbVO$_4$ can be ascribed to the easy saturation of the magnetization along the easy-direction under low magnetic fields (1.3 T at 2 K). However, when increasing temperature, the resulting thermal fluctuations enhances the required field to completely align the Tb$^{3+}$ magnetic moments and maximising accordingly the MCE. This explains why high magnetic fields are needed to obtain the table-like profile of $\Delta S_T$ (T) shown in Figure 6. Finally, the structural transformation occurring around 33 K which is associated with a cooperative Jahn Teller transition in TbVO$_4$ also helps enlarging its working cryogenic temperature range.

It is worth noting that the obtained maximum isothermal entropy change under a magnetic field variation of 7 T applied along the easy axis is practically equal to the theorical limit given approximately by $\Delta S_{T\ Limt} = R*\ln(2) = 21$ J/ kg·K ($R$ is the universal gas constant) since the ground state is a (quasi) doublet at zero field. TheTbVO$_4$ compound keeps this value approximately constant up to roughly 34 K because of its large magnetic moment and pseudo-Ising character [26, 32-38]. In contrast to the easy axis, the TbVO$_4$ single crystal unveils moderate isothermal entropy changes along the c-axis because of the presented gigantic magnetocrystalline anisotropy. Under a magnetic field changing from 0 to 2 T along the hard axis, the associated maximum isothermal entropy change is only about 1.6 J/kg K.

These findings show that the TbVO$_4$ could be an excellent magnetocaloric candidate for application in cryo-magnetic coolers. However, the strong MCE shown by TbVO$_4$ is also of great importance from a fundamental point of view since its presence would markedly distort the conclusions in relation with performed studies under intense magnetic fields. This could be the case of recently reported works investigating the effect of high magnetic fields on the Jahn-Teller transitions in RVO$_4$ orthovanadates [27-29]. For example, in order to directly study the Jahn-Teller transition, Detlefs et al. [28] have carried out x-ray diffraction patterns on TbVO$_4$ at the cryogenic temperature of 7.5 K under magnetic fields varying from 0 up to 30 T. In a similar context, Hazei et al [30] have investigated the critical field associated with the destruction of quadrupole ordering (increasing symmetry) in TbVO$_4$ under the application of strong pulsed magnetic fields that can reach 50 T in the temperature range between 1.5 and 30 K. However, the application of such intense magnetic fields at temperatures below 30 K will markedly shift the compound temperature far away from its initial temperature due to the magnetocaloric effect. Consequently, the real temperature of measured samples under strong magnetic fields are much higher than those mentioned in related works. This would result in spurious interpretation of obtained results. Keeping in mind this point, the accurate measurement of MCE prior to high magnetic field experiments is necessary for a better assessment of resulting data.

In addition to large thermal effects that can be induced by changing magnetic fields along the a axis, the TbVO$_4$ single crystals reveal a strong anisotropy of the magnetocaloric effect between the easy and hard directions as shown in Figures 3 and 4. When subjected to an applied magnetic field variation of 5 T along the easy axis, the TbVO$_4$ exhibits an isothermal entropy change 10 times larger than its equivalent along the hard axis. Consequently, strong MCEs could be generated by simply rotating TbVO$_4$ single crystals between their easy and hard magnetization orientations in constant magnetic fields without the need to the conventional magnetization-demagnetization method. Such anisotropic MCEs [22, 49-51] would enable us to design a new generation of compact, efficient and simplified rotary magnetic cooling systems that could be dedicated to the liquefaction of hydrogen and helium [22, 23]. Similarly to the standard MCE, the isothermal entropy change associated with the rotation of TbVO$_4$ single crystals between two different crystallographic directions can be also determined from magnetic isotherms. Considering the magnetic field initially applied along the c direction, the obtained rotating magnetocaloric effect (RMCE) in term of $\Delta S_T$ when spinning the crystals between the hard and easy axes by an angle of 90° in a constant magnetic field H is simply given by [1, 14, 16-18]:

$$\Delta S_{c-a}(T, H) = \Delta S(T, H//easy) - \Delta S(T, H//hard) \qquad (6)$$

where $\Delta S_T$ (H//easy) and $\Delta S_T$ (H//hard) are the isothermal entropy changes corresponding to the variation of the magnetic field along the easy and hard directions, respectively. The temperature dependence of $\Delta S_{c-a}$ under some selected constant magnetic fields is plotted in Figure 8. Since it is difficult to maintain the hard direction of TbVO$_4$ parallel to applied high magnetic fields because of the resulting torques in the sample, our study of the RMCE was limited to magnetic fields going from 0 up to 5 T. As shown, large rotating isothermal entropy changes which are similar to the conventional ones are also obtained at low temperatures. The rotation of TbVO$_4$ crystals in a constant magnetic field of 2 T allows to generate a maximum isothermal entropy change that exceeds 18 J/kg K close to 3 K. More interestingly, as shown in Figure 8 the $\Delta S_{c-a}$ varies slightly in the temperature range below 30 K under high magnetic fields being similar to the standard $\Delta S_T$ features (Fig. 6). For example, the maximum obtained $\Delta S_{c-a}$ at 6.5 K under 5 T (19 J/kg K) decreases by about 20 % when the temperature is increased up to 30 K. Similarly to the conventional isothermal entropy change, $\Delta S_{c-a}$ is expected to remain approximately constant over a wide temperature range under sufficiently high magnetic fields. Such a broadening of $\Delta S_{c-a}$ (T) leads to a large rotating refrigerant capacity that reaches about 460 J/kg (for 5 T), being much larger than the RC reported in some of the best oxides known for their large RMCE at low temperature [14]. This result unveils that the gaseous hydrogen and helium can be also liquefied by using TbVO$_4$ crystals in more efficient, compact, and simplified cryocoolers working with RMCEs [1, 14, 16-18, 49-51].

As demonstrated in this paper, the TbVO$_4$ vanadates reveals fascinating magnetic and magnetocaloric properties with potential implementation in low-temperature magnetic refrigerators. These features are strongly coupled to the crystal electrical field and the crystallographic symmetry. For a good understanding of the magnetism and the magnetocaloric effect including anisotropic MCEs in TbVO$_4$, it is worthy to recall how the crystal field acts on the RVO$_4$ electronic ground state.

In this family of materials, the JT transition is associated with small displacements of ligand ions changing accordingly the impact of the crystal field on rare-earth magnetic ions. Such effect usually takes place in the case where accidental orbital degeneracy or near-degeneracy exist defining then the energies of the low-lying states. According to pioneer theoretical works and optical spectroscopy [24, 25, 32, 52], it clearly seems that in the RVO$_4$ vanadates, the crystal field at the R$^{3+}$ ions gives rise to large magnetic quadrupole moments in the xy plane at the tetragonal-high temperature phase. This results in the near-degeneracy of the lowest electronic state thanks to the fourfold symmetry about the c-axis as underlined by Elliot et al [24]. Consequently, the tetragonal crystal field of D$_{2d}$ symmetry in the TbVO$_4$ system induces four closely spaced low-lying levels with a large separation from the next states [52]. Such a situation leads to a singlet ($\Gamma_1$)-doublet($\Gamma_5$)-singlet ($\Gamma_3$) equally separated by roughly 9 cm$^{-1}$. In the DyVO$_4$ system, the Dy$^{3+}$ which is a Kramers ion gives a pair of doublets spaced each to other by 9 cm$^{-1}$ [24, 25, 32, 52].

The analysis of stain modes in RVO$_4$ unveils B$_{1g}$ and B$_{2g}$ symmetries to be responsible for the lowering of the crystallographic symmetry from tetragonal to orthorhombic structure. In this way, it is worthy to recall that in TbVO$_4$ the Tb$^{3+}$ ions have two equivalent rare earth sites at the tetragonal phase with D$_{2d}$ symmetry while in the orthorhombic low-temperature phase, the Tb$^{3+}$ sites remain equivalents with D$_2$ symmetry [24, 53]. On the other hand, the investigation of crystal field parameters leads to a quartet ground state constituted of $|J_x\rangle, |J_y\rangle, |J_{-x}\rangle$ and $|J_{-y}\rangle$ [25, 54, 55, 56]. The latter which exhibit flat pancakes forms [25] intersect along the c axis giving rise to magnetic (or electric) quadrupoles. However, for the TbVO$_4$, the sign of crystal field parameters shows that above states point along the x' and y' axes with the angle xox' being π/4. The lowering of its crystallographic symmetry which is associated with energy lowering (JT effect) at roughly 33 K occurs via a B$_{2g}$ distortion giving at the crystal field symmetry of D$_2$ two near-degenerate doublets separated by approximately 50 cm$^{-1}$ as reported in Gehring et al [25] and Bleaney et al [26]. In fact, at temperatures below 33 K, the Γ$_3$ singlet transforms into a Γ$_1$ singlet with a raised energy level placed at 51.3(5) cm$^{-1}$ while the Γ$_5$ doublets splits into two singlets under JT effect. The first (Γ$_2$) is situated at the energy level 47.0(5) cm$^{-1}$ while the other becomes Γ$_4$ with reduced energy just above Γ$_1$ [26]. At the orthorhombic phase, both Γ$_1$ and Γ$_4$ are non-magnetic singlets [37]. The established doublets are split under the effect of external magnetic field with g-factor depending on crystal axes leading to a strong magnetism following the ab-plane and gigantic magneto-crystalline anisotropy in TbVO$_4$ [38]. For the latter, the g-value was found by Gehring et al [38] to be 16.3 along [010] direction or equivalent and 0 for other directions at 4.2 K. Such resulting magnetic behaviors greatly impact the magnetocaloric properties and particularly the magnitude of the rotating MCE as demonstrated in the present paper. Regarding other RVO$_4$ vanadates, the crystal field in HoVO$_4$ results in a ground state that can be represented by a non-magnetic singlet combined with an excited state doublet. The latter is placed at the crystal field state 20 cm$^{-1}$ and splits below 20 K with a separation of 2 cm$^{-1}$ [57, 58] making from the HoVO$_4$ a typical van Vleck paramagnet. This would explain its relatively low magnetocaloric properties as reported in Dey et al [31]. For DyVO$_4$, the measurement of g-factors by Ranon [59] gives 9.903 along the a axis and 1.104 along the c axis. This make Dy$^{3+}$ magnetic moments order in an "Ising-like" magnetic shape [32]. In the case of ErVO$_4$, Er$^{3+}$ is a Karmers doublet where the g-factors along the a and c axes are also reported by Ranon [59] to be 7.085 and 3.544, respectively.

To learn more about the crystal field excitation of Tb$^{3+}$ in TbVO$_4$ and to support our analysis, we have measured the infrared transmission spectra of TbVO$_4$ as a function of temperature with a resolution of 1 cm$^{-1}$. Figure 9 shows the recorded transmission spectra of Tb$^{3+}$(4f$^8$) $^2$F$_6$→ $^2$F$_4$ CF transitions in TbVO$_4$ at 5 K, 14 K, 30 K, 40 K and 200 K for the oscillating electric field parallel to the ab-plane. Below 33 K, all the frequencies of absorption bands are split and shifted to higher energy. For example, the excitation at 3657 cm$^{-1}$ splits clearly at 30 K. The component at higher energy continues to shift with decreasing temperature while the component at lower frequency loses intensity and almost vanishes at 5 K. Such a behavior is a sign of a replica absorption band of the thermally excited levels of the $^2F_6$ ground state. The magnitude of this splitting is around 48 cm$^{-1}$. The absorption bands at 3488 cm$^{-1}$, 3411 cm$^{-1}$ and 3322 cm$^{-1}$ behave similarly and with the same splitting magnitude. This magnitude of 48 cm$^{-1}$ is exactly equal to the splitting of the $\Gamma_5$ doublets into $\Gamma_2$ and $\Gamma_4$ under the JT effect as reported in Ref. 26. Our findings are in fair agreement with previous studies and point out to the role of Tb$^{3+}$ crystal field scheme in the gigantic magnetocrystalline anisotropy shown by the TbVO$_4$ compound [38]. The behaviour of Tb$^{3+}$ crystal-field excitations in TbVO$_4$ under magnetic fields will be the subject of further future investigations by our team.

## IV. Conclusions

To sum up, we have investigated the magnetic and magnetocaloric properties of TbVO$_4$ single crystals. Raman scattering data confirms the lowering of crystallographic symmetry from the high-temperature tetragonal phase to the low-temperature orthorhombic structure at 33 K which is associated with a Jahn-Teller distortion involving the Tb$^{3+}$ orbitales. The newly established symmetry results in a strong magnetocrystalline anisotropy with hard and easy magnetization directions along the c and a axes. Since the V$^{5+}$ valence shell is empty, the shown magnetism is fully contributed from the Tb$^{3+}$ magnetic moments that tend to order at temperatures below 4 K. On the other hand, the application of low magnetic fields reachable by permanent magnets along the easy axis enables us to totally align the Tb$^{3+}$ magnetic moments giving rise to a strong magnetocaloric effect. The latter remains approximately constant over a wide cryogenic temperature range under sufficiently high magnetic field leading to a record refrigerant capacity. Additionally, such a behavior is highly appropriate for cryo-magnetic coolers using the more efficient Ericsson and AMR thermodynamic cycles. More importantly, the established gigantic magnetic anisotropy between the easy and hard orientations at the temperature range below 33 K permits also strong thermal effects that can be induced simply by rotating TbVO$_4$ crystals between both directions in constant magnetic fields. Such RMCEs would open the way for the implementation of TbVO$_4$ in more efficient and simplified cryogenic magnetic refrigerators. To avoid the economical and practical drawbacks usually associated with the growth of single crystals, powders of the TbVO$_4$ orthovanadate can be textured (oriented) under magnetic fields. Finally, for a better assessment of performed experiments on RVO$_4$ orthovanadates under intense magnetic fields, the resulting MCE must be measured and considered prior to such investigations.

**Acknowledgments:**

The authors thank M. Castonguay, S. Pelletier and B. Rivard for technical support. We acknowledge the financial support from NSERC (Canada), FQRNT (Québec), CFI, CIFAR, Canada First Research Excellence Fund (Apogée Canada), Université de Sherbrooke, and the International University of Rabat. D.Z.D and J.-Y.J. acknowledge funding by the Ministry of Science and Technology, Taiwan, grant numbers MOST 106-2112-M-009-013-MY3 and MOST 108-2811-M-009-538 and by the Bulgarian National Fund (BNF) project DCOST 01/2. M. B would like to thank M. Bikerouin for his help to design the $TbVO_4$ crystallographic structure.

**Author Contributions**

M. B. conceived the work, performed magnetic measurements, analyzed data, prepared figures, and wrote the paper. S. M. performed Raman scattering and infrared crystal field transition measurements and revised the paper. D. Z. D. and J. Y. J. prepared the needed single crystals. S. J and P. F. revised the paper.

**Competing Interests:** The authors declare that they have no competing interests

**Figure captions**

**Figure 1**: Crystallographic structure of $TbVO_4$

**Figure 2:** (a) Raman spectra of $TbVO_4$ at some selected temperatures. (b) Raman spectra of $TbVO_4$ at the frequency range between 800 and 850 $cm^{-1}$.

**Figure 3:** ZFC and FC thermomagnetic curves of the $TbVO_4$ single crystal under low magnetic fields applied along the easy and hard axes. Inset: 0.1 T- ZFC and FC thermomagnetic curves following the easy direction in the temperature range between 31 and 37 K.

**Figure 4:** (a) ZFC reverse magnetic susceptibility as a function of temperature of the $TbVO_4$ single crystal along the easy and hard axes. (b) Magnetic isotherms of the $TbVO_4$ single crystal at 2 K along the easy and hard axes. Inset: Magnetization in the ab plane along the a and b axes at 2 K.

**Figure 5:** Isothermal magnetization curves of the single crystal $TbVO_4$ for (a) H//a and (b) H//c. The increments of temperature are 1 K from 3 to 10 K, 2 K from 10 to 20 K and 4 K above 20 K.

**Figure 6:** Temperature dependence of the isothermal entropy change in TbVO$_4$ under different magnetic field variations for (a) H//a and (b) H//c.

**Figure 7:** Refrigerant capacity of TbVO$_4$ single crystal compared to some relevant cryo-magnetocaloric oxides [14, 33, 47, 48] in the magnetic field change of 7 T.

**Figure 8:** Isothermal entropy changes related to the rotation of TbVO$_4$ between the easy and hard axes by 90 ° in different constant magnetic fields, with magnetic field initially parallel to the c-axis.

**Figure 9:** Transmission spectra of Tb$^{3+}$ $^2F_6 \rightarrow$ $^2F_4$ crystal field transitions in TbVO$_4$ at 5 K, 14 K, 30 K, 40 K and 200 K.

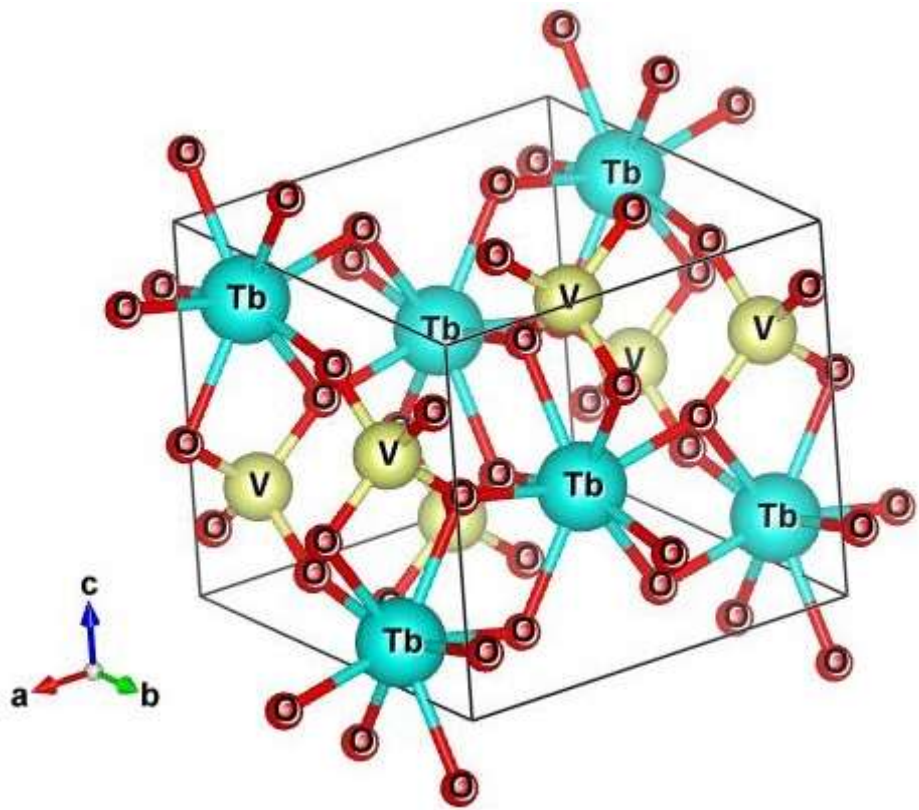

**Figure 1**

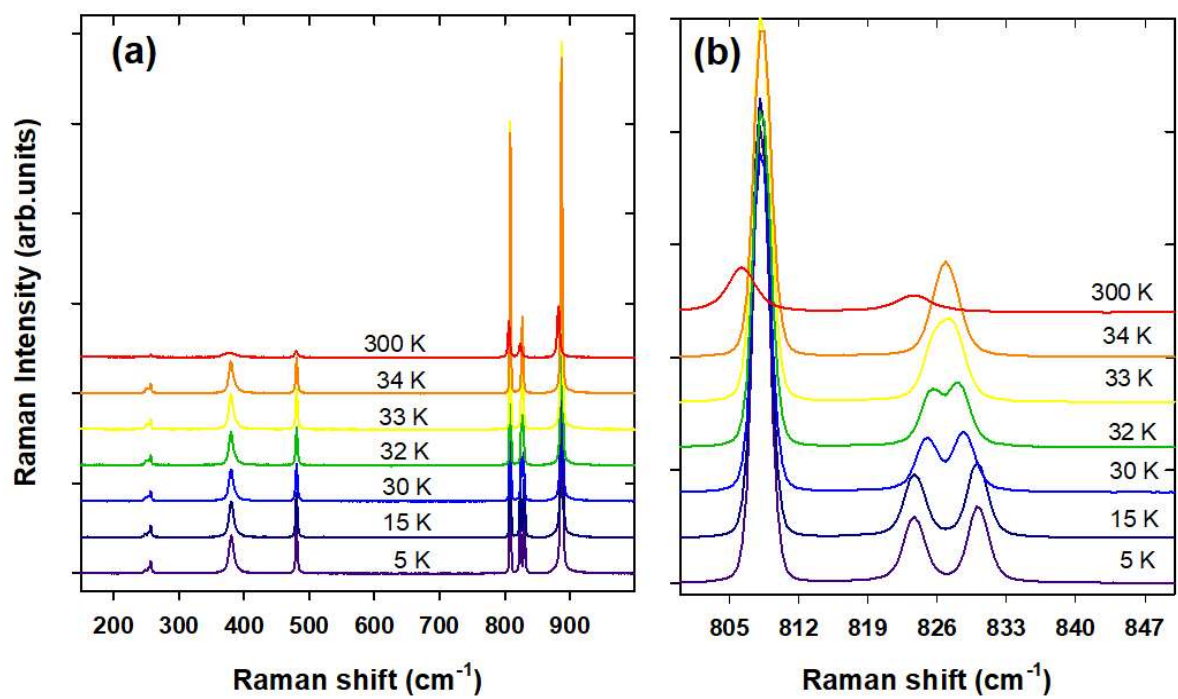

**Figure 2**

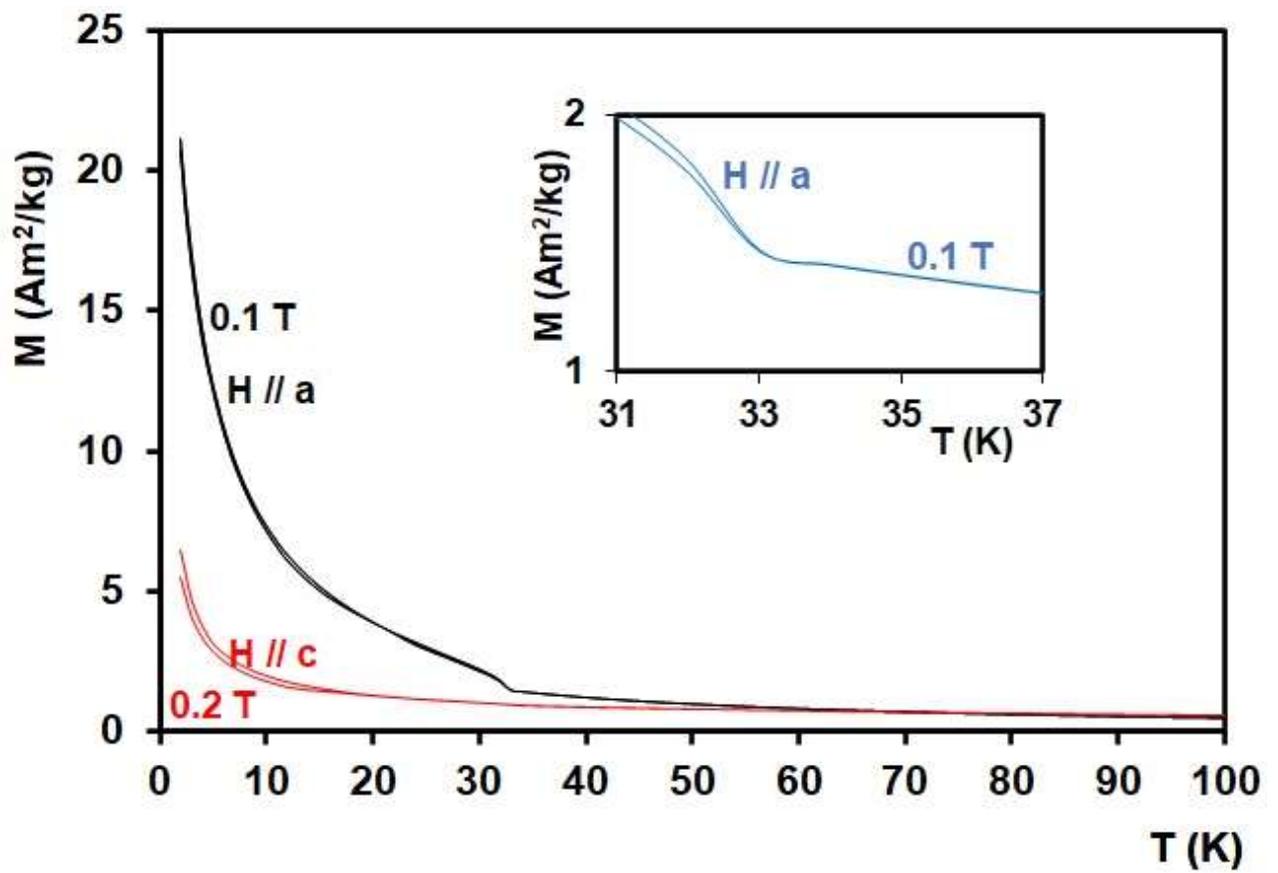

Figure 3

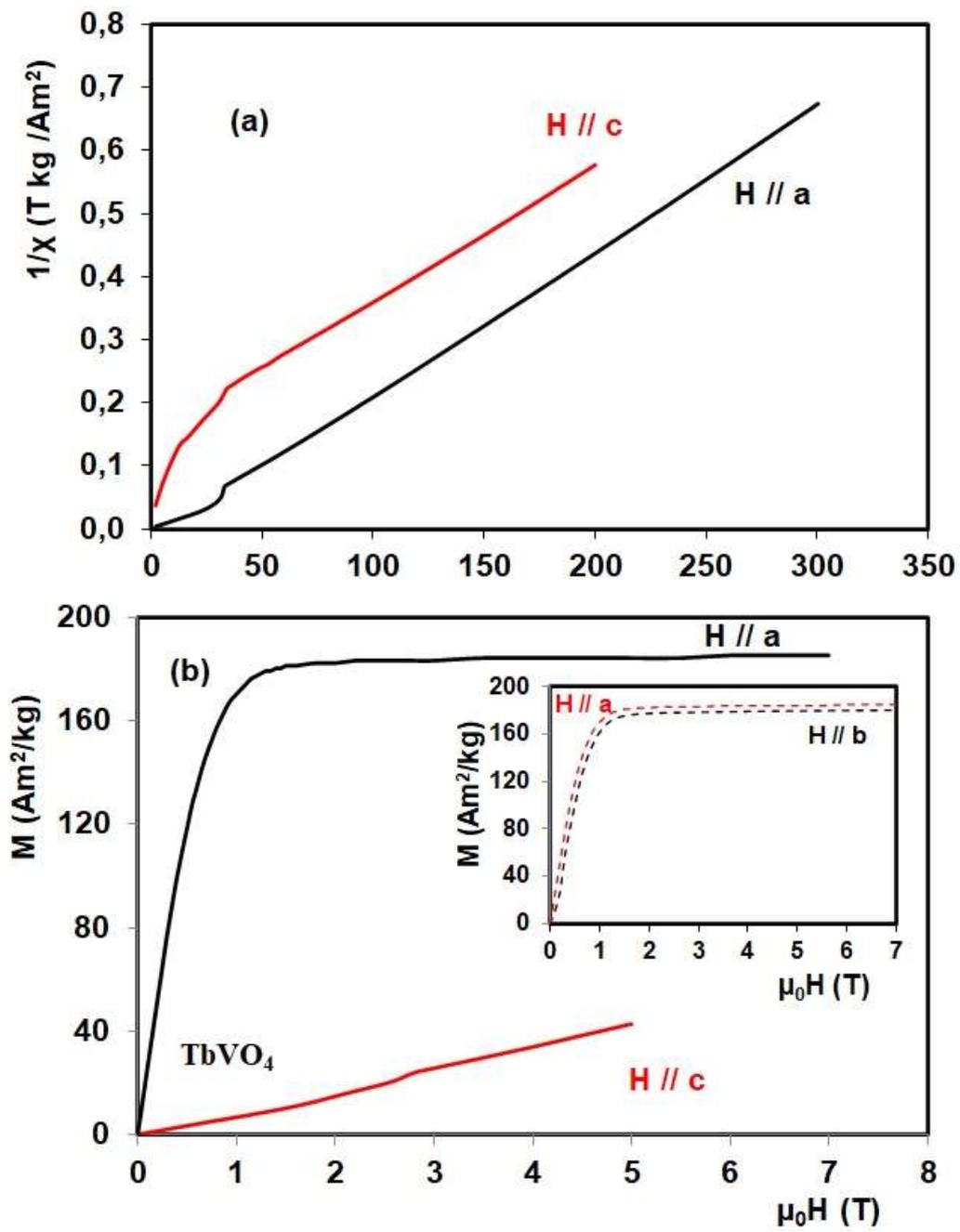

**Figure 4**

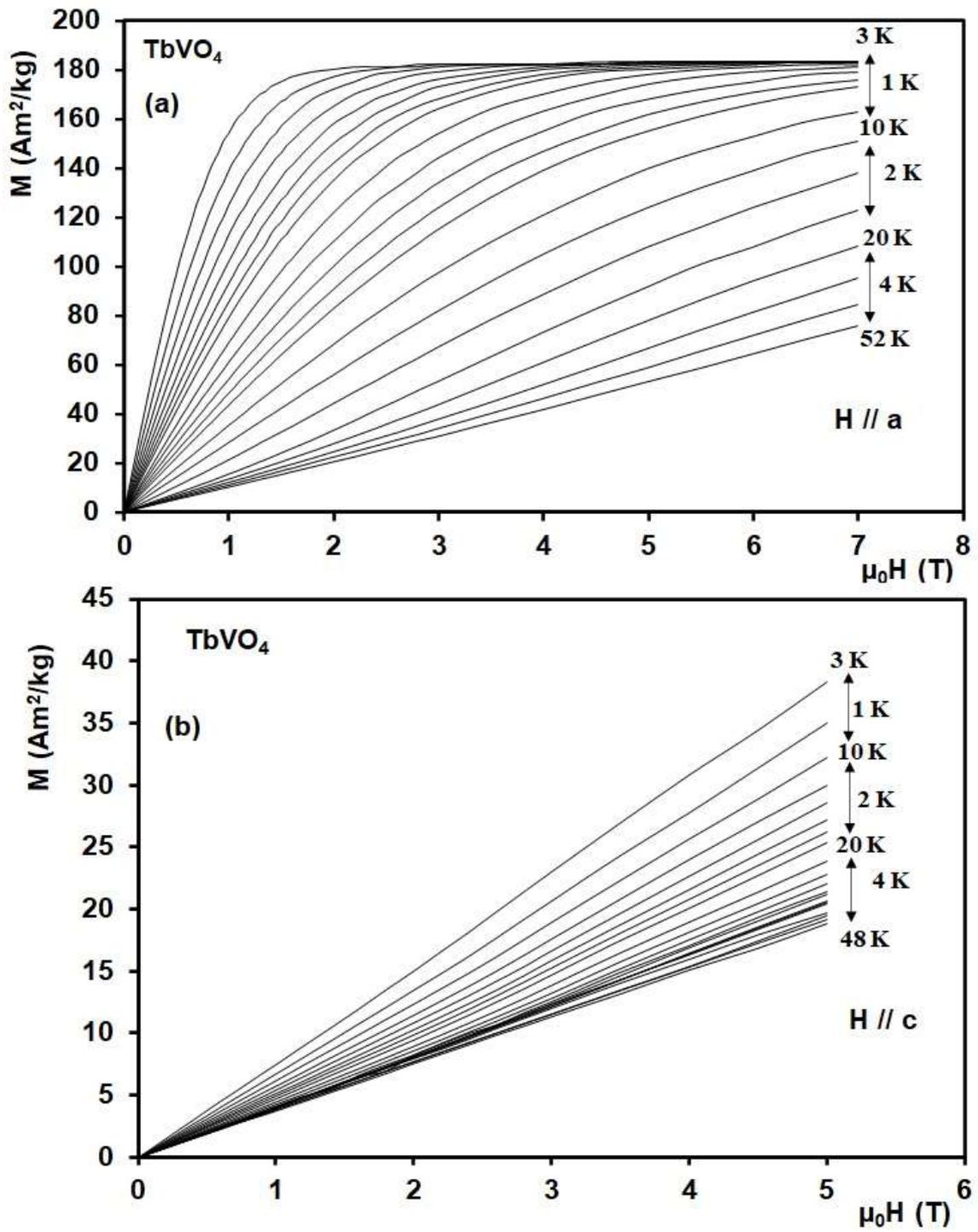

**Figure 5**

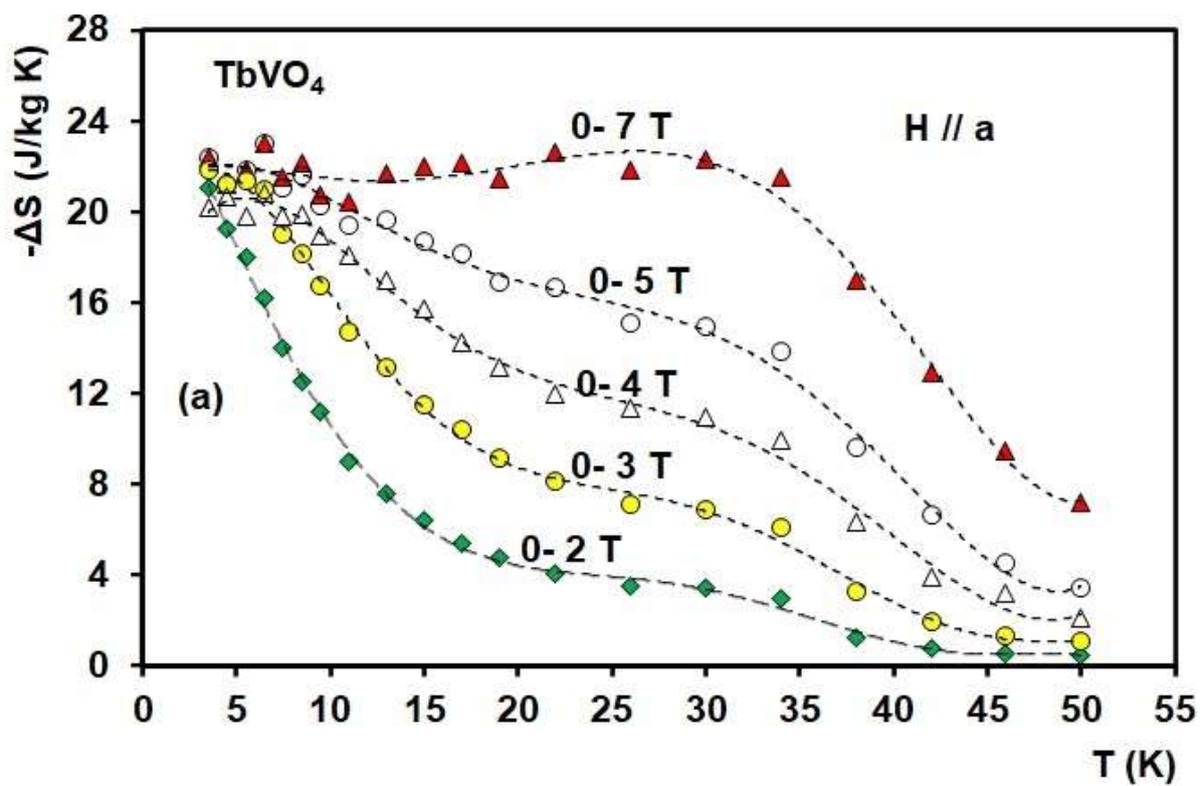

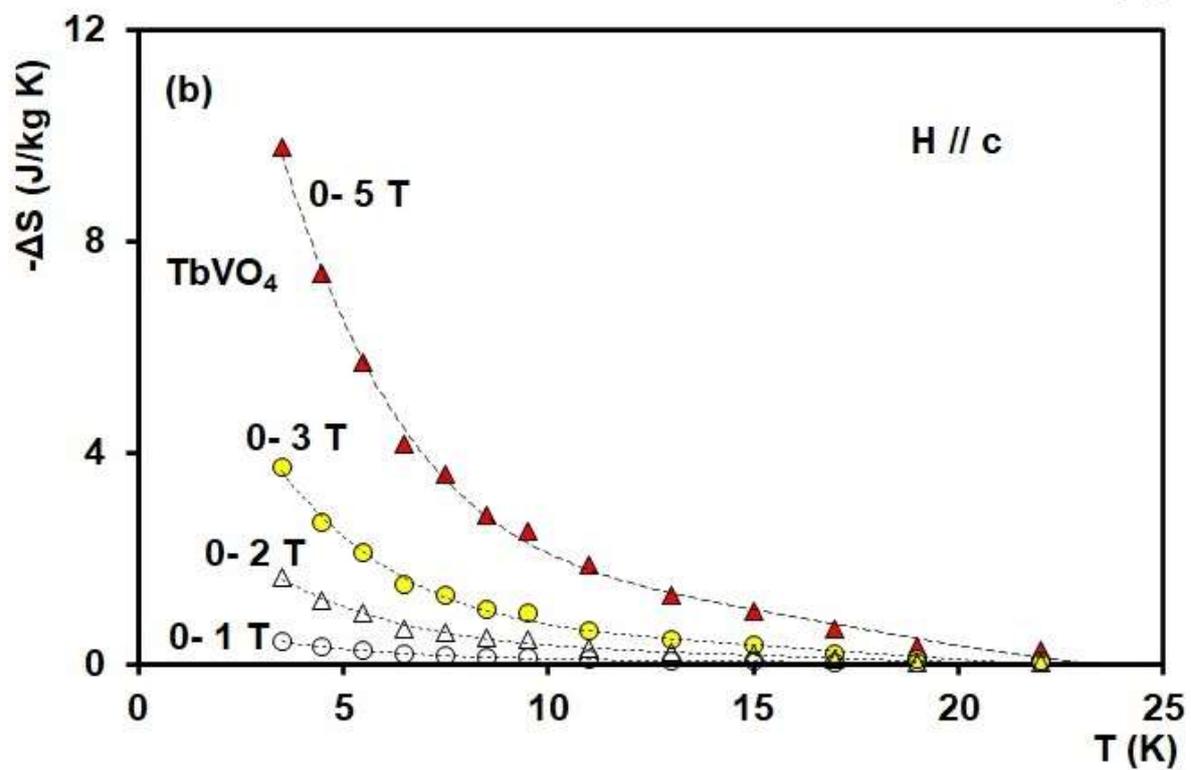

**Figure 6**

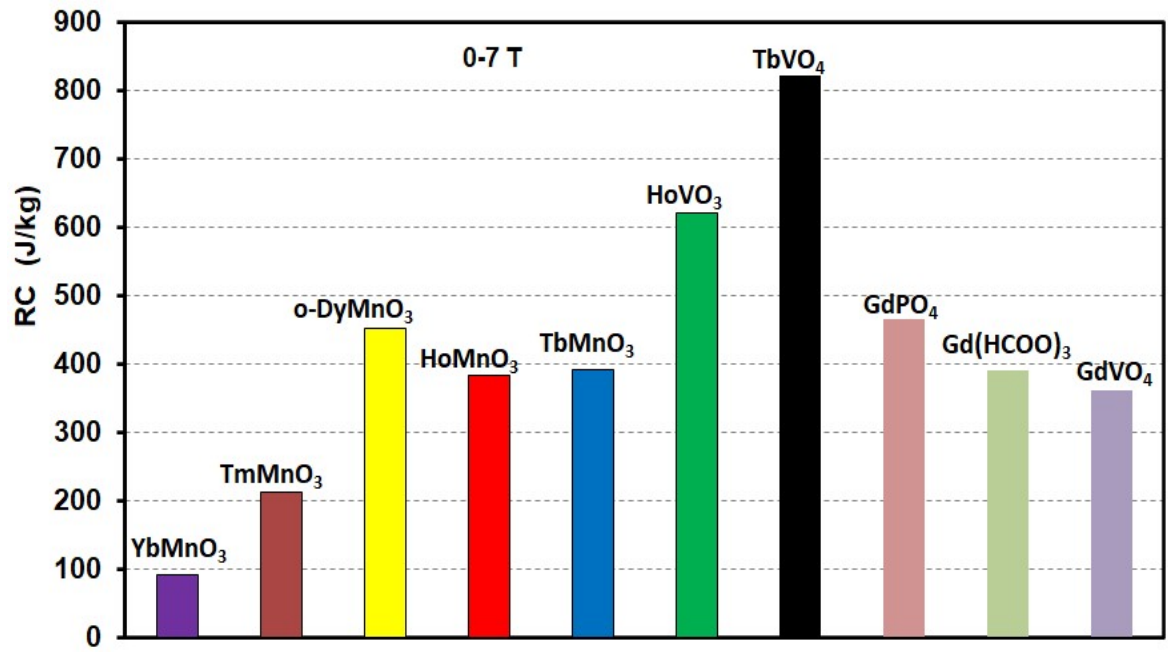

**Figure 7**

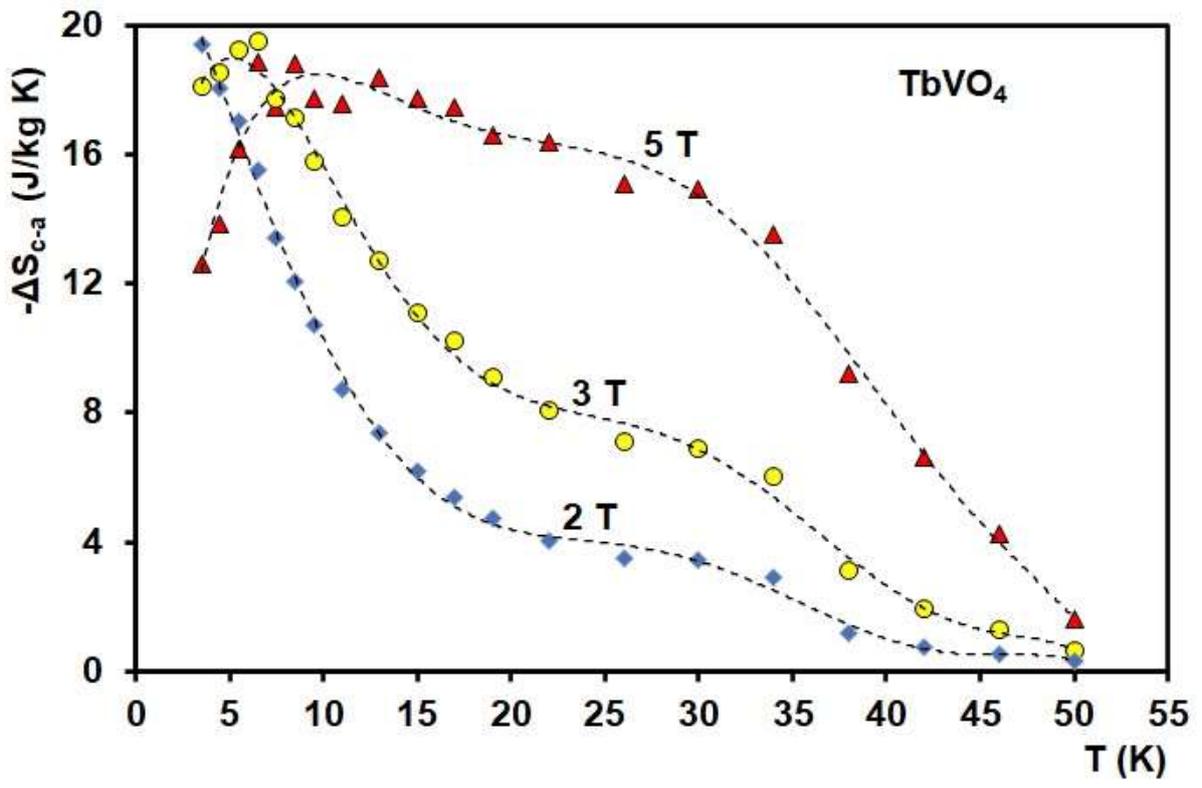

**Figure 8**

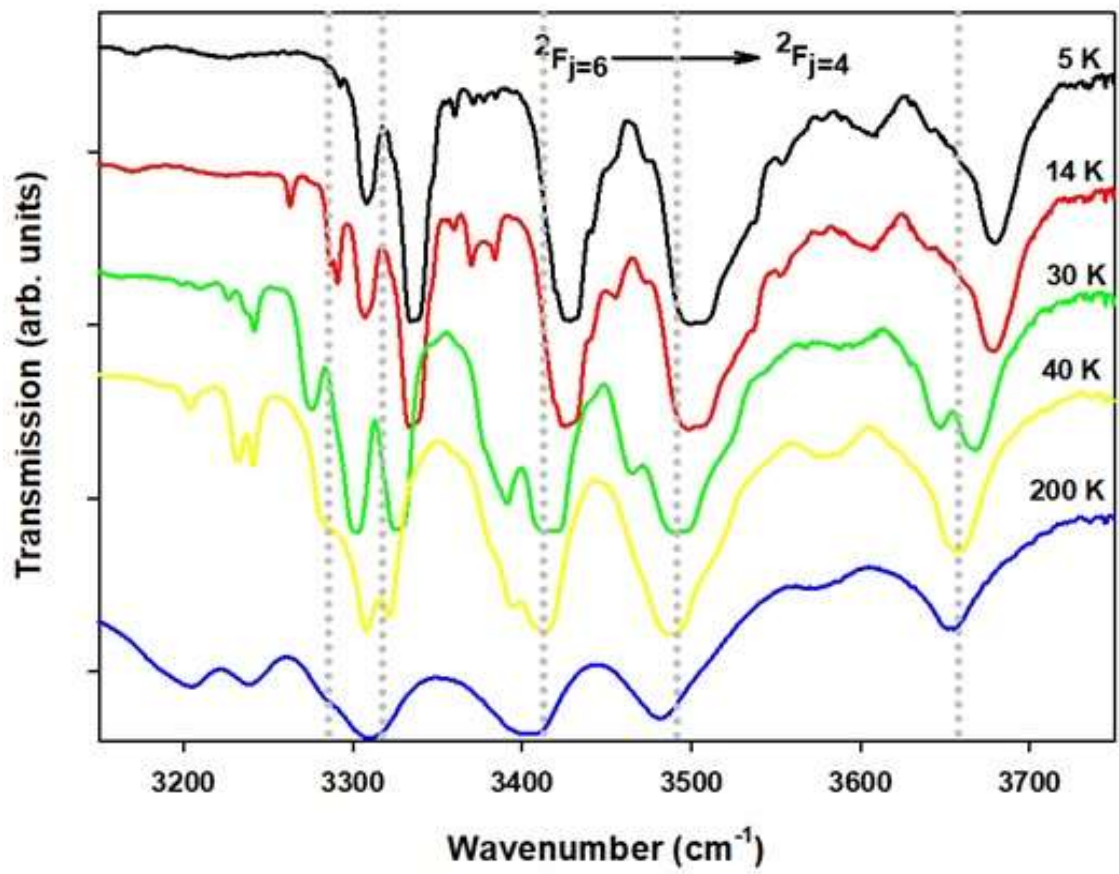

Figure 9